\documentclass[useAMS]{mn2e}
\usepackage{psfig}

\usepackage{psfig}
\usepackage{times}

\title[Optical afterglow luminosities in the {\it Swift} epoch]
{Optical afterglow luminosities in the {\it Swift} epoch:
confirming clustering and bimodality
}

\author[Nardini, Ghisellini \& Ghirlanda]
{M. Nardini $^{1}$\thanks{E--mail: nardini@sissa.it},
G. Ghisellini $^{2}$ and 
G. Ghirlanda $^{2}$\\
$^{1}$SISSA--ISAS, Via Beirut 2-4, 34314, Trieste, Italy\\
$^{2}$Osservatorio Astronomico di Brera, via Bianchi 46, I--23807
Merate, Italy.
}
\begin{document}

\maketitle


\begin{abstract}

We show that Gamma Ray Bursts (GRBs) of known redshift and rest frame optical 
extinction detected by the {\it Swift} satellite fully confirm earlier results
concerning the distribution of the optical afterglow luminosity at 12 hours
after trigger (rest frame time).
This distribution is bimodal and relatively narrow, especially for the 
high luminosity branch. 
This is intriguing, given that {\it Swift} GRBs have, on average,
a redshift larger than pre--{\it Swift} ones, and is
unexpected in the common scenario explaining the GRB afterglow.
We investigate if the observed distribution can be the result of selection effects 
affecting a unimodal parent luminosity distribution, and find that either
the distribution is intrinsically bimodal, or most (60 per cent) of the bursts are
absorbed by a substantial amount of grey dust.
In both cases we suggest that most dark bursts should belong to the underluminous 
optical family.

\end{abstract}

\section{Introduction}

After 3 years since the launch of the {\it Swift} satellite (Gehrels et al. 2004),
the number of long GRBs with known redshift strongly increased. 
The optical multiband follow up of these events allowed also the analysis of the 
spectral energy distribution for a large fraction of their optical afterglows. 

In Nardini et al. (2006a) we analysed the optical (in the $R$ band) luminosity
distribution after 12h (rest frame time) of all the 24 pre--{\it Swift}  long
GRBs with known redshift and a published estimate of the host galaxy
absorption $A_V^{host}$.  Most of them (i.e. 21/24) had optical luminosities
$\log{L_{\nu_R}}$ that lie in a very narrow range with mean value 
$\langle \log{L_{\nu_R}}\rangle =30.65$ 
(monochromatic luminosities in units of erg s$^{-1}$ Hz$^{-1}$), 
with a dispersion $\sigma=0.28$.  The remaining 3 events
were about 15 times fainter than the main group. 
The clustering of the observed optical afterglow luminosities
for most GRBs, and the hint of a bimodality, with a few underluminous events 
(3.6 $\sigma$ smaller luminosities) have been found  
independently by Liang \& Zhang (2006) and confirmed also in 
Nardini et al. (2006b) who considered a small number of GRBs detected by 
{\it Swift} (but with their intrinsic optical extinction still unpublished).

In Nardini, Ghisellini \& Ghirlanda (2008, hereafter NGG08) we tested 
whether the observed clustering and bimodality of the optical afterglow
luminosities could be due to possible intervening selection effects. 
Our simulations showed that
the observed distribution could be obtained either if there are 
indeed two intrinsically separated GRB families,
or if there is a large amount (more than 1.5 magnitudes) of
unrecognised achromatic (grey) absorption affecting most 
(but not all) sources.
The existence of an underluminous family, or the presence of grey dust, could
explain why a sizeable fraction of GRBs are optically dark.  The
observed underluminous events should represent the ``tip of the iceberg'' of a
much more populated optically fainter family that includes the undetected dark
GRBs.  
J\'ohannesson Bj\"ornsson \& Gudmundsson (2007) found that the
observed bimodality of the optical luminosity distribution can be reproduced
in the standard fireball model only assuming a bimodality in the
intrinsic model parameters (i.e. two families with different 
mean isotropic kinetic energy $E_0$). 
However,  the latter result seems not to agree with the X--ray data.

In this work we consider all GRBs detected after the launch of {\it Swift}
with known redshift and $A_V^{host}$.
The obtained optical luminosity distribution confirms the results obtained in
Nardini et al. (2006a) for what concerns both the clustering of the luminous
family and the bimodal nature of the distribution. 
We then repeat the exercise of NGG08 on this larger sample of events to test 
whether the clustering and bimodal luminosity distribution is intrinsic or due 
to some selection effect.  
To this aim we collected all upper limits of the optically dark GRBs detected
in the same {\it Swift} epoch. 

While we were completing our study, Kann et al. (2008)
  confirmed the bimodality of the intrinsic optical luminosities
  considering a slightly differently selected sample of events, and noted also 
a strong similarity between the pre--{\it Swift} and the {\it Swift} 
luminosity distributions.

 We use $H_0=70$ km s$^{-1}$ Mpc$^{-1}$ and $\Omega_\Lambda=0.7$, $\Omega_{\rm M}=0.3$.

\section{The {\it Swift} GRB sample}

As of December 2007, there are 
85 long GRBs detected in the last 3 years with a spectroscopic redshift
determination.  
  For 45 of them the light curve is sampled well enough to
  allow a good determination of $L_{\nu_R}$  at 12 h rest frame.  
In 29 cases there is a published estimate of the host galaxy dust
absorption $A_V^{host}$.  
These 29 events are then included in our sample
together with the pre--{\it Swift} bursts (24 GRBs selected
with the same criteria).  
We also add two pre--{\it Swift} GRBs whose $A_V^{host}$ has been 
recently published: 
GRB 040924 ($\log{L_{\nu_R}}=28.85$) with $A_V^{host}=0.16$ given by 
Kann, Klose \& Zeh (2006); and 
GRB 041006 ($\log{L_{\nu_R}}=29.38$) with $A_V^{host}=0.14$ (Misra et al. 2005).
The total sample then includes 55 GRBs.

\begin{figure}
\vskip -0.7 true cm
\hskip -9 true cm
\psfig{figure=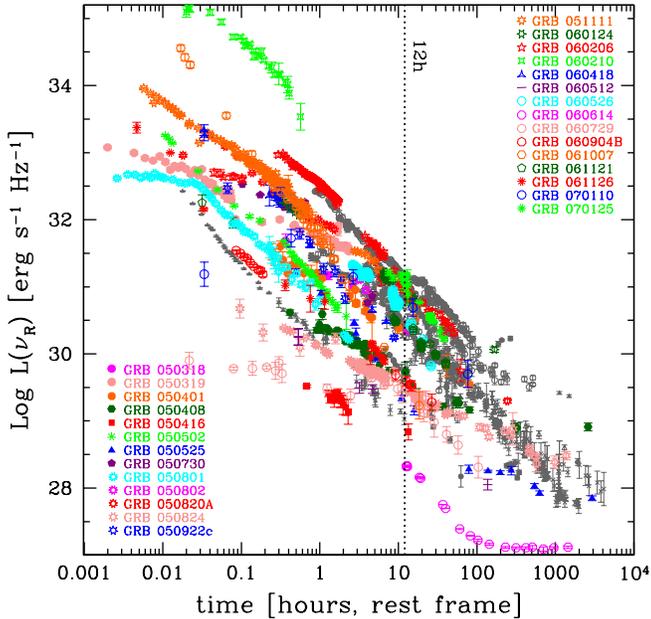,angle=0,width=10cm}
\vskip -0.8 true cm
\caption{
  Optical monochromatic luminosity $\log{L_{\nu_R}}$ light curves of long GRBs
  with a published $A_V^{host}$ estimate.  Time is in the rest frame of the
  source.  Grey dots represent the pre--{\it Swift} sample from Nardini et al.
  (2006a).  Coloured points correspond to the {\it Swift} GRBs analysed in
  this work.  All these {\it Swift} bursts are labelled.  The vertical line is
  at 12 h.  }
\label{lc}
\end{figure} 

Fig. \ref{lc} shows the rest frame light curves of all the {\it Swift} events
superposed to the pre--{\it Swift} sample (grey symbols).  
Note that the new sample shows a much denser photometry coverage at early times 
(i.e. at $t$\textless 1h after trigger).  
These light curves show that at early times the
behaviour is different in different bursts, to become more ``regular" and
similar at later times when it becomes also similar to the optical decay
typical of the pre--{\it Swift} epoch.  

We compute the luminosity of each burst at a common time of 12h after trigger because 
i) it gives a good representation of the late time afterglow behaviour
without showing the peculiar features sometimes appearing in the first hour;
ii) it is usually before the jet break time;
iii) it can be easily compared with the pre--{\it Swift} results.  
A different rest frame time choice in the interval between 2 hours and 2 days
would not affect significantly our results.

\begin{figure}
\vskip -0.7 true cm
\hskip -9 true cm
\psfig{figure=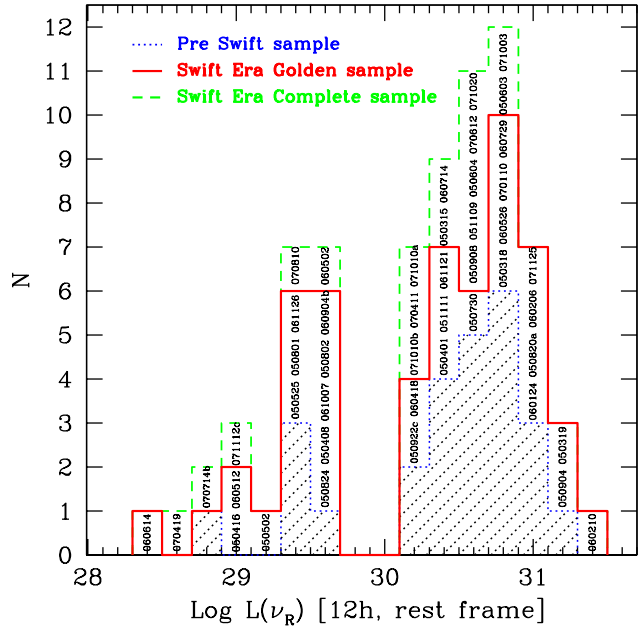,angle=0,width=10cm}
\vskip -0.8 true cm
\caption{
  Optical luminosity distribution at 12h rest frame time.
  The dashed area shows the pre--{\it Swift} distribution
  (Nardini et al. 2006a) with the addition of 2 new GRBs. 
  The continuum red
  line represents the sum of the pre--{\it Swift} bursts and the 
  {\it Swift} GRBs with published $A_V^{host}$.
  The dashed green line includes those {\it Swift} bursts 
  with no published $A_V^{host}$.}     
\label{istotutti}
\end{figure}

In order to evaluate the monochromatic luminosity at 12h, we interpolated the
photometric $R$ band points around this time.
As mentioned, we did not extrapolate data taken before 1h from the trigger,
not to be biased by the possible peculiar behaviour of the very early afterglow.

\subsection{Comparison with the pre--{\it Swift} sample}

Fig. \ref{istotutti} shows the distribution of the luminosities at 12h
for the {\it Swift} GRBs superposed to the pre--{\it Swift} ones.  We can see
that the {\it Swift} bursts fully confirm the pre--{\it Swift} results.
Both the clustering of the brighter luminosities and
the separation between the luminous and the subluminous families are strongly
confirmed.  The pre--{\it Swift} high luminosity family (Nardini et al. 2006a)
was well fitted by a log--normal distribution with a mean value of $\langle
\log{L_{\nu_R}^{12h}}\rangle =30.65$ and a dispersion $\sigma=0.28$.  The
entire sample ({\it Swift} and pre--{\it Swift} GRBs) now has 
$\langle \log{L_{\nu_R}^{12h}}\rangle=30.71$ and $\sigma=0.31$ (these values becomes
$\langle \log{L_{\nu_R}^{12h}}\rangle=30.65$ and $\sigma=0.31$ if we also
consider the events without a host extinction estimate).  
A Kolmogorov Smirnov (KS) test yields a probability $P\approx$ 28\% that
the pre--{\it Swift} and the {\it Swift} distributions come from the same parent 
population. This result does not rule out the hypothesis that we are observing
  the same parent burst population, even if the {\it Swift} sample has a mean redshift 
($\langle z \rangle = 2.0$)
larger than the pre--{\it Swift} one ($\langle z \rangle = 1.4$).  
There are several {\it Swift} GRBs belonging to the underluminous family,
yet no GRB falls into the luminosity gap between the two families.
This strengthens the possible existence of a bimodal luminosity 
distribution (see below).
The ratio between the underluminous and the luminous pre--{\it Swift} GRBs 
is 5/21 (with the addition of GRB 040924 and GRB 041006, it was
3/21 in Nardini et al. 2006a). 
For {\it Swift} GRBs this ratio becomes closer to unity (i.e. 12/17 for 
GRBs with $A_V^{host}$ known and 16/29 if we include {\it Swift} bursts with no 
published $A_V^{host}$). 
The improved optical telescopes capabilities (see below) allowed the detection of
a number of very faint events with $\log{L_{\nu_R}^{12h}}<29.0$ and increased
the number of detectable members of the faint family.

\section{Telescope Selection Function (TSF) for {\it Swift} bursts}

In order to analyse the possible selection effects affecting the optical
observations in the {\it Swift} epoch, we consider the optical upper limits
obtained when the burst, observed and localised by the X--ray
  telescope [XRT] onboard {\it Swift}, is observed at optical
  wavelengths but not detected.
The main difference with
respect to the pre--{\it Swift} epoch is that now the optical afterglow can be
followed even 100 s after the trigger (in some cases even at shorter times),
thanks to UVOT, the optical--UV monitor onboard {\it Swift}, and by 
ground based robotic telescopes.

In NGG08 we created the distribution of the deepest $R$ band upper limits of
dark GRBs at the observed time of 12 h.  
These limits were derived by extrapolating, at 12h, all upper limits
for each burst, assuming a time decay $f(t)\propto t^{-\alpha}$ with
$\alpha=1$  (typical value of the optical decay at these time scales (Zhang 2007)).
Then we choose the deepest value.
These were corrected for the Galactic absorption
along the line of sight using Schlegel et al.  (1998).  
This correction accounts for the limitation in the telescope
sensitivity affecting the obtained upper limit.
The obtained distribution can be considered as the probability,
for each burst, to be observed at 12 hours with a telescope 
(and an exposure time) reaching a given magnitude limit.
We call it the ``Telescope Selection Function" (TSF).

In the pre--{\it Swift} epoch it was believed that all optical afterglows
had a similar decay, while we now know that the situation is more complex.
The flat shape of a large number of very early optical afterglows does not
allow to use the simple assumption of a single power law decay lasting from
few seconds to days after the trigger.  Using very early photometric
upper limits in order to extrapolate an upper limit at 12h  by assuming a   
$f(t)\propto t^{-1}$ decay would lead to a strong overestimate of the latter 
because that choice does not account for the flatter early time  
light curve shape, leading to a too severe constraint on the afterglow 
flux at later times. 
To be conservative we decided not to use the
upper limits obtained before 1h after the trigger in order to determine the
deepest upper limit at 12h to build the TSF.  
This choice allows us also to better compare the {\it Swift} results 
with the ones obtained in NGG08. 
We have chosen 1h as the minimum time because the optical light curve,
which can be flat at earlier times, seems to recover the pre--{\it
Swift} behaviour after this time. Note that 1 h  
  is of the order of the (observer frame) time $T_a$ found
  by Willingale et al. (2007) for the ``flat--steep" transition
  of the X--ray light curves. 
  Note that Gendre, Galli \& Bo\"er (2008),
  in their analysis of X--ray afterglow luminosities, also
  adopted the choice, similar to ours, of considering data only after $T_a$.

We analysed all the optical limiting magnitudes of the 146 long GRBs without
optical detection.  
Of these, 20 were not observed in the optical.  
For the remaining 126 bursts,  
we found 74 GRBs with at least one useful $R$ band upper limit.  
For the other bursts the available data were taken before 1h.  
We did not use the unfiltered observation.

\begin{figure}
\vskip -0.5 true cm
\hskip -8.5 true cm
\psfig{figure=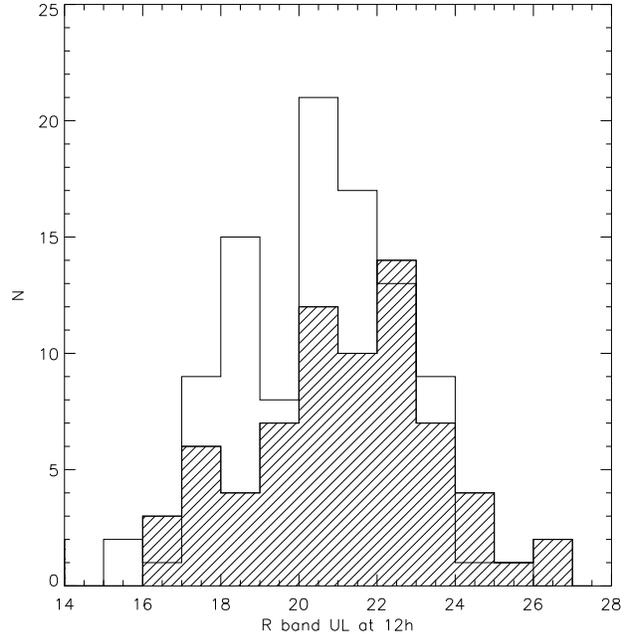,angle=0,width=9cm}
\caption{
Distribution of the deepest $R$ band upper limits (greater than 15) of 
all dark GRBs, evaluated at 12 hours. The dashed area represents the 
{\it Swift} dark GRB sample while the empty area represents the pre--{\it Swift} 
sample considered in NGG08. 
All upper limits are corrected for  the Galactic absorption. 
The plotted distribution is the TSF quoted in the text. }     
\label{tsf}
\end{figure}

If we compare the obtained TSF with the pre--{\it Swift} one shown in 
NGG08 (see Fig. \ref{tsf}) through a KS test, we find that the two 
distributions are different at the $\sim$2$\sigma$ level 
(the KS null hypothesis probability is 5\%). 
We note that, even if we do not consider the very early upper limits, 
this new distribution appears slightly deeper than the previous one.
Indeed, the mean value of the upper limits at 12h is 0.9 magnitude deeper than
the pre--{\it Swift} TSF.  This difference decreases to
0.63 if we do not take into account the events with a very weak upper limits
(i.e. $R$ band upper limit \textless 14).  The fraction of very deep upper
limits ($R>24$) moves from 2\% to 9.5\%.

\section{Testing the bimodality}

In the previous sections we showed that the {\it Swift} bursts confirm both
the clustering and the bimodality of the observed optical luminosity
distribution.  
NGG08, considering only pre--{\it Swift} bursts, proved that
these result cannot be explained invoking a selection effect affecting an
intrinsically unimodal broad luminosity function and assuming 
``normal" (i.e. chromatic) absorption. 
Kann et al.  (2008) claimed
that the results of NGG08 cannot be directly applied to 
the {\it Swift} sample because both the assumed TSF and luminosity 
distributions were built considering telescopes with a lower capability 
of observing faint events with respect to the {\it Swift} epoch.  

The method proposed in NGG08 basically tries to reproduce the
luminosity distribution of Fig. \ref{istotutti} 
considering the optical selection effects introduced by the TSF on an
assumed intrinsic GRB luminosity function.  We simulated 30000 GRB optical
afterglows assuming a redshift distribution (traced by the cosmic star
formation rate described by Porciani \& Madau 2001), an intrinsic luminosity
function, a host galaxy dust absorption distribution and the probability
distribution for each burst to be observed with a telescope with a given
sensitivity.  
The latter distribution can be well
represented by the TSF obtained above (Fig. \ref{tsf}).  
In order to compare the simulated result with the observed 
distribution plotted in Fig. \ref{istotutti} that includes both the 
pre--{\it Swift} and {\it Swift} GRBs, we created a combined TSF that 
includes all the upper limits contained in the two TSFs.  
  This combined TSF takes into account the number of GRBs
  observed in the optical in the pre--{\it Swift} and {\it Swift} epochs.
  Considering both the detected and undetected optical afterglows, these
  are 156 and 249, respectively.
We assume that the GRBs of the two samples
intrinsically belong to the same distribution and that the differences in the
observed distributions are just due to the change of the observing conditions.

The simulation selects all the events whose observable (i.e. corrected for
cosmological effects and for the host galaxy dust absorption) flux is
larger than the upper limit of the assigned telescope.  
Note that the Galactic
absorption is already considered in the upper limit definition and that
all GRBs with $z>5$ are considered undetectable because of the Ly$\alpha$
absorption in the $R$ band.  
The optical luminosity distribution of the resulting simulated events 
can be compared with the real one shown in Fig. \ref{istotutti}.  
For a more detailed discussion of this method see NGG08.

\subsection{Comparison with the observed distribution}

The number of the events in Fig. \ref{istotutti}, especially the ones
belonging the underluminous family, strongly increased with respect to the
pre--{\it Swift} sample.

  As in the pre--{\it Swift} sample case we cannot
  use the standard two tailed Kolmogorov Smirnov (KS) test because of its
  weak sensitivity for the tails of the distribution and when 
  comparing an unimodal with
  a bimodal distribution with a similar median. In these cases the the KS test
  strongly overestimates the probability for the distributions to be generated
  from the same parent one. Similarly to what we did in NGG08, we compare the
  simulated and the observed distributions using the Cash (1979) test dividing the
  observed luminosity range into 11 bins.
  We reject a model under test 
  if the $C$ factor is larger than 9.2 ($P_{rej}>99\%$) even if in some cases we
  find $C$ values smaller than 4.6 ($P_{rej}=1-P_C>90\%$).
As done in NGG08 we tested a Gaussian, Top Hat and Power Law ($N_L\propto
L^{-\delta}$) unimodal luminosity function defined over the same luminosity
range of the comparison distribution of Fig. \ref{istotutti}.  
The largest observed luminosity gives a strong constraint because 
more luminous events would have been easily seen.  
The low luminosity threshold is less constrained
from Fig. \ref{istotutti} and it is more affected by observational limits.  For
a very modest host galaxy dust absorption ($A_V^{host}<1$) we obtain:
\begin{itemize} 
\item Top Hat with $28.4<\log{L_{\nu_R}}<31.3$: $P_C=1.3\times 10^{-3}$\%;
\item Power law in the same range and $\delta=2$: $P_C=8.2\times 10^{-4}$\%;
\item Gaussian with $\mu=29.9$, $\sigma=0.7$: $P_C< 10^{-5}$\%.
\end{itemize}
Our results show that we cannot reproduce the observed distribution of
Fig. \ref{istotutti} with a unimodal luminosity distribution of GRBs.  This
result is almost independent from the assumed dust distribution, if it is
standard.  We used a Small Magellanic Cloud extinction curve because it seems
more appropriate to represent the GRB afterglow host galaxy extinction (Schady
et al. 2007, Kann et al. 2006), but we obtain similar results using the Milky
Way and the Large Magellanic Cloud extinction curves.

Much better agreement is obtained either assuming an intrinsic bimodal
luminosity function or assigning to most of the events an additional
achromatic dust absorption.  Note that this ``grey dust" absorption is
elusive, and cannot be estimated by the usual technique used to find
$A_V^{host}$, namely assuming an intrinsic power law shape of the optical
spectrum.  A grey dust extinction has been invoked earlier for explaining some
puzzling GRB spectral energy distributions (e.g.  Perna \& Lazzati 2002,
Stratta et al. 2005, Perley et al. 2007).

The best results for the different tested luminosity distributions have been
obtained assuming, together with the ``grey dust" extinction, a moderate
standard reddening modelled with a simple top hat $A_V^{host}$
distribution between 0 and 1.8 magnitudes.  
The best matches between the simulated and the observed distribution are:
\begin{itemize}
\item 
Gaussian $\sigma=0.30$ $\mu=30.69$, $A_{grey}=1.6$: $P_C=0.9$\%: 
\item 
Top Hat with  $30.1<\log{L_{\nu_R}}<31.3$, $A_{grey}=1.6$: $P_C=11.6$\%
\item 
Power law in the same range and $\delta=2$, $A_{grey}=1.6$: $P=11.1$\%
\end{itemize}
The increased number of underluminous observed events with respect to the
pre--{\it Swift} sample 
gives more information about the shape of the fainter family
distribution. 
The Gaussian luminosity case is ruled out
because it does not well represent the fainter events distribution (which is
now more populated). 
The simple assumption of
adding a strong (about 1.6 magnitudes) achromatic absorption to about 60\%
of the events is still producing acceptable results ($P_{C}\approx$ 10-12\%).
In future, with improved statistics, we will have to
better characterise either the fainter family luminosity function (in the
case of an intrinsically bimodal function) or the achromatic absorption
distribution.
 
Liang \& Zhang (2006) and Kann et al. (2008) noted that the mean redshift of
the fainter family is smaller than the more luminous ones. 
For the present sample the mean redshift of the faint group is 
$\langle z\rangle =1.17$ vs $\langle z\rangle =2.4$ of the luminous family.
  
Also in our simulated sample the observable
GRBs belonging to the fainter family have a mean redshift of
$\langle {z}_{faint}\rangle \approx 1.74$ vs $\langle {z}_{bright}\rangle 
\approx 2.17$\footnote{These results are obtained in the power--law bimodal 
scenario but similar results are obtained in all the other cases}, 
even if their intrinsic redshift distribution is the same. 
Our simulated faint events seem to be located
at larger redshifts with respect to the observed ones while the 
simulated and observed $\langle {z}_{bright}\rangle $ are comparable.

In our simulations about 1/3 of the $z<5$ events are observable and
most of the undetectable ones are members of the low luminosity family.
This suggests that dark GBRs preferentially are optically underluminous GRBs.

\section{Discussion and conclusions}

We have shown that {\it Swift} bursts confirm the distribution
of the luminosities of the optical afterglows observed at a fixed time (12h)
in the rest frame of the source: there are two families, 
both contained in a narrow luminosity range, and with a gap
between the two.
The ratio of the averaged luminosities of the two families is about 25 
(i.e. $\mu_{faint}$=29.3, $\mu_{bright}$=30.7).
We proved that this observed 
dichotomy is not due to some simple intervening observational
selection effects, 
but it must corresponds to an intrinsic bimodality either of the 
afterglow luminosity function itself or of the distribution of 
the absorption, with half of the burst affected only by moderate 
``normal" (i.e. chromatic)
extinction, and the other half dimmed by a further
1.5--2 mag of ``grey" dust absorption.
The first possibility suggests a dichotomy of the intrinsic properties
of the burst, while the second suggests a dichotomy of the properties
of the GRB environment.
We cannot (yet) distinguish between the two possibilities,
but an increase of the number of GRBs (say, twice as many as we have now,
with redshift, well monitored
optical afterglow and estimate of the ``normal", chromatic, host extinction) 
will make it possible to well constrain either the slope of the luminosity 
function or the shape of the grey absorption distribution.
Other crucial information will come from the analysis 
of the optical to X--ray afterglow evolution.
This will tell us if they belong or not to the same component
(see e.g. Panaitescu 2007), with important consequences on the sometimes puzzling
connection between these bands 
(e.g. Stratta et al. 2005, Urata et al. 2007, Troja et al. 2007, li,
li \& Wei 2008, Perley et al. 2008) and the nature of the host galaxy
dust absorption.      

Also for the X--ray luminosities there is some evidence of 
clustering and dichotomy, but there is no simple connection 
between this bimodality and the one observed in optical 
(Gendre \& Bo\"er 2005; Gendre, Galli \& Bo\"er 2008). 
Also the energetics of the prompt emission seem unrelated to the
afterglow luminosities:
the events with the lower bolometric isotropic energy 
$E_{\gamma, iso}$ are also members of the optically fainter family 
(e.g. GRB 050416, GRB 060614, GRB 030329), but there are
optically faint events with high  $E_{\gamma, iso}$ (e.g. GRB 061007,  
GRB 061126). 
We therefore confirm what found in Nardini et al. 2006a, i.e. there is
no evident correlation between  
$E_{\gamma, iso}$ and the optical luminosities (see also Kann et al. 2008). 

The observed optical luminosity distribution has no convincing explanation yet, 
but it can shed new light for understanding the existence 
of the optically dark long GRBs. 
During the past 3 years after the launch of {\it Swift}, there have been 
167 GRBs with at least one optical afterglow detection. 
If we define dark GRBs all the events observed, but not detected
in the optical, and for which there is an optical upper limit,
we find 126 events in these 3 years, i.e. about 40\% of all long GRBs
in the {\it Swift} epoch. 
In our simulations about 2/3 of the $z<5$ events are undetectable 
and most of them are members of the low luminosity family. 
This overestimate of the number of the simulated dark burst is probably due to 
the assumption (made for simplicity) that the faint and the bright 
families have the same shape (even if with different normalisations).
With this caveat in mind, we suggest that dark GBRs preferentially 
are optically underluminous GRBs.

\section{Acknowledgments}
We would like to thank a 2005 PRIN-INAF grant for funding and 
Annalisa Celotti for useful discussions. 
We would also like to thank the referee, Bruce Gendre, for his 
constructive comments.

\section{Appendix: Sources of data}
GRB 050318: Still et al. (2005);
GRB 050319: Tagliaferri et al. (2006);
GRB 050401: Watson et al. (2006);
GRB 050408: de Ugarte Postigo et al. (2007);
GRB 050416: Holland et al. (2007);
GRB 050502: Yost et al., (2006), Kann et al. (2008);
GRB 050525: Blustin et al. (2006), Kann et al. (2008);
GRB 050730: Starling et al. (2005), Pandey et al. (2006), Kann et
al. (2008);
GRB 050801: Kann et al. (2008);
GRB 050802: Oates et al. (2007), Kann et al. (2008);
GRB 050820A: Cenko et al. (2006), Kann et al. (2008);
GRB 050821: Sollerman et al. (2007), Schady et al. (2007), Kann et al. (2008);
GRB 050922C: Kann et al. (2008);
GRB 051111: Butler et al. (2006), Schady et al. (2007);
GRB 060124: Misra et al. (2007);
GRB 060206: Thone et al. (2008),  Kann et al. (2008);
GRB 060210: Curran et al. (2007);
GRB 060418: Ellison et al. (2006), Molinari et al. (2007), Schady et
al. (2007), Kann et al. (2008); 
GRB 060512: Schady et al. (2007);
GRB 060526: Kann et al. (2008); 
GRB 060614: Mangano et al. (2007);
GRB 060729: Grupe et al. (2007);
GRB 060904B: Kann et al. (2008); 
GRB 061007: Mundell et al. (2007), Kann et al. (2008); 
GRB 061121: Page at al. (2007); 
GRB 061126: Perley et al. (2008), Kann et al. (2008);
GRB 070110: Troja et al. (2007);
GRB 070125: Kann et al. (2008).

\end{document}